\documentclass[11pt]{article}
\usepackage{scicite}
\usepackage{times}
\usepackage{graphicx}

\newenvironment{sciabstract}{%
\begin{quote} \bf}
{\end{quote}}

\title{Continuous-Wave Frequency Upconversion with a Molecular Optomechanical Nanocavity}

\author
{Wen Chen,$^{1}$ Philippe Roelli,$^{1\dagger}$, Huatian Hu,$^{2}$, Sachin Verlekar,$^{1}$\\ Sakthi Priya Amirtharaj,$^{1}$ Angela I. Barreda, $^{3}$ Tobias J. Kippenberg,$^{1}$\\ Miroslavna Kovylina,$^{4}$ Ewold Verhagen,$^{5}$  Alejandro Mart\'{i}nez,$^{4}$\\ Christophe Galland$^{1\ast}$ 
\\
\normalsize{$^{1}$Institute of Physics, Ecole Polytechnique F\'ed\'erale de Lausanne (EPFL)}\\
\normalsize{CH-1015 Lausanne, Switzerland}\\
\normalsize{$^{2}$Hubei Key Laboratory of Optical Information and Pattern Recognition}\\
\normalsize{Wuhan Institute of Technology, Wuhan 430205, China}\\
\normalsize{$^{3}$Institute of Applied Physics, Abbe Center of Photonics}\\
\normalsize{Friedrich Schiller University Jena, Albert-Einstein-Str. 15, 07745 Jena, Germany}\\
\normalsize{$^{4}$Nanophotonics Technology Center, Universitat
Politècnica de València}\\
\normalsize{Camino de Vera s/n, 46022 Valencia, Spain}\\
\normalsize{$^{\dagger}$Present address: Nano-optics group, CIC nanoGUNE BRTA, San Sebastián, Spain}\\
\normalsize{$^{5}$Center for Nanophotonics, AMOLF, Science Park 104, 1098 XG Amsterdam, Netherlands}\\
\\
\normalsize{$^\ast$To whom correspondence should be addressed; E-mail:  chris.galland@epfl.ch}
}

\date{}

\begin{document} 

\baselineskip14pt

\maketitle 
\begin{sciabstract}
  Frequency upconversion is a cornerstone of electromagnetic signal processing, analysis and detection. 
  It is used to transfer energy and information from one frequency domain to another where transmission, modulation or detection is technically easier or more efficient.
  Optomechanical transduction is emerging as a flexible approach to coherent frequency upconversion; it has been successfully demonstrated for conversion from radio- and microwaves (kHz to GHz) to optical fields. 
  Nevertheless, optomechanical transduction of multi-THz and mid-infrared signals remains an open challenge.
  Here, we utilize molecular cavity optomechanics to demonstrate upconversion of sub-microwatt continuous-wave signals at $\sim$32~THz into the visible domain at ambient conditions. 
  The device consists in a plasmonic nanocavity hosting a small number of molecules.
  The incoming field resonantly drives a collective molecular vibration, which imprints an optomechanical modulation on a visible pump laser and results in Stokes and anti-Stokes upconverted Raman sidebands with sub-natural linewidth, indicating a coherent process.
  The nanocavity offers 13 orders of magnitude enhancement of upconversion efficiency per molecule compared to free space, with a measured phonon-to-photon internal conversion efficiency larger than $10^{-4}$ per milliwatt of pump power.
  Our results establish a flexible paradigm for optomechanical frequency conversion using molecular oscillators coupled to plasmonic nanocavities, whose vibrational and electromagnetic properties can be tailored at will using chemical engineering and nanofabrication.
\end{sciabstract}
\paragraph*{Introduction}

Our ability to generate, detect and analyse electromagnetic signals spanning the full spectrum from radiofrequency up to visible light or even X-rays governs technological progress in areas ranging from information processing, telecommunication networks, material characterisation, broadband imaging and molecular sensing. 
In this broad context, frequency conversion plays an essential role by allowing detection and modulation of signals that are in a frequency band where suitable technologies are not widely accessible.
A prominent example is that of electromagnetic waves with frequencies in the range from $\sim$1 to $\sim$100 THz  \cite{tonouchi2007}, which we will refer to here as infrared (IR) waves in the broad sense.
They find applications ranging from homeland security and molecular analysis of gases, chemicals and biological tissues \cite{bruyne2018,bai2021}, to thermal imaging and non-destructive material inspection \cite{ciampa2018}, to astronomical surveys \cite{roellig2020}.
Yet, they represent a technological frontier for signal detection and processing, with a scarce offer of commercial devices in comparison with telecommunication or visible bands.
Spectroscopy of IR signals is typically performed using tunable laser sources \cite{tittel2003} or broadband sources coupled to a Fourier-transform IR (FTIR) spectrometer.
These methods require bulky apparatuses or expensive coherent sources, and are typically limited to slow acquisition rates.
Moreover, they employ IR detectors \cite{rogalski2019} that are more noisy than visible and near-infrared (VIS/NIR) detectors, motivating recent efforts to perform FTIR using NIR light with a nonlinear interferometer \cite{lindner2021}. 

As a powerful and flexible approach to analyze long-wavelength electromagnetic signals, frequency upconversion is being intensely studied for imaging and spectroscopy applications \cite{barh2019}. 
The main motivation for frequency upconversion in the context of spectroscopy is to leverage inexpensive, fast and low-noise detector technologies available in the VIS/NIR domain \cite{roelli2020}.
This method has the additional advantage of allowing, in principle, coherent mapping of a quantum state from one frequency to another, with broad applications in classical and quantum information technologies. It offers a way to connect different information processing nodes over long distances via optical fibers \cite{radnaev2010, zaske2012, higginbotham2018, lauk2020, lambert2020}. 
Frequency upconversion of IR signals can be accomplished by three-wave mixing in a bulk crystal with a large effective second order nonlinearity. 
Delicate phase matching, large pump powers and cm-long crystals are typically needed to reach high efficiencies \cite{karstad2005,tidemand-lichtenberg2016,junaid2018,tseng2018,pedersen2020}, even though plasmonic gap modes have recently been leveraged to dramatically enhance nonlinear effects \cite{nielsen2017,ummethala2019,salamin2019,gordon2019}.
A second-order nonlinearity is also present at interfaces, where inversion symmetry is broken. 
This fact has been broadly leveraged in ultrafast nonlinear spectroscopy to probe the properties and dynamics of molecular layers on various surfaces \cite{shen1989,tian2014} 
with applications extending to biologically relevant membranes \cite{roke2012}.
In this context, resonant excitation of vibrational modes is responsible for molecule-specific signatures in the three-wave mixing signal, which offers deep insights into molecular structure, dynamics, and molecule-surface interactions \cite{shen1989,tian2014}.
However, such techniques require substantial peak powers only accessible with femto- or picosecond pulses and have therefore remained spectroscopic tools used by an expert community.

Optomechanical cavities have recently emerged as promising candidates to realise quantum coherent frequency conversion \cite{chu2020,lauk2020,lambert2020}.
In this implementation, the signal of interest resonantly drives a mechanical oscillator, itself parametrically coupled to a laser-driven optical cavity, which results in modulation sidebands at the sum- and difference-frequencies (called anti-Stokes and Stokes sidebands, respectively).
This approach offers a number of advantages, such as the resonant enhancement of nonlinear response at the mechanical frequency and the parametric enhancement of conversion efficiency with intracavity pump power.
It is also highly versatile and has been demonstrated with mechanical resonance frequencies ranging from kHz \cite{bagci2014,moaddelhaghighi2018} to GHz \cite{bochmann2013,andrews2014,vainsencher2016,balram2016,vanlaer2018,forsch2020,mirhosseini2020}. 
In a different approach, modulations on THz waves have been read out optically through a MHz-frequency mechanical resonator \cite{belacel2017}.
Molecular oscillators constitute a new frontier in cavity optomechanics \cite{roelli2016,schmidt2017}, as they enable to reach multi-THz resonant frequencies and room-temperature quantum coherent operation \cite{velez2020}. 
Moreover, they can be coupled to plasmonic nanocavities with deep sub-wavelength mode volumes, thereby enabling optomechanical coupling rates in excess of 1~THz \cite{benz2016a}.
Despite these promising features, devices based on molecular cavity optomechanics have yet to be demonstrated.



Here, we experimentally demonstrate the upconversion of a 32~THz (9.3~$\mu$m wavelength) continuous wave IR signal into the visible domain using a sub-wavelength molecular optomechanical cavity at ambient conditions. 
Our upconversion scheme operates with micro-Watt level continuous wave signal and pump beams, which is a disruptive departure from alternative methods, and allows high-resolution spectroscopy of the IR signal thanks to the coherent nature of optomechanical transduction.
This novel regime of operation is achieved by coupling a monolayer of molecular oscillators to a doubly resonant plasmonic gap nanocavity, which supports deep-sub-wavelength mode volumes and simulated field enhancement factors in excess of 500 and 100 at IR and VIS frequencies, respectively, from which an overall enhancement of upconversion efficiency per molecule by more than 13 orders of magnitude compared to free space is predicted and experimentally confirmed.
In our proof-of-principle experiment, internal optomechanical upconversion efficiency per unit of pump power are on the order of $10^{-4}$ per milliwatt with an estimated 10\% outcoupling efficiency, while the IR photon-to-vibration conversion efficiency is on the order of $10^{-5}$, corresponding to end-to-end conversion efficiencies above $10^{-12}$ for 10~$\mu$W pump power; we present several routes to further improve these figures. 


\begin{figure}
    \centering
    \includegraphics[scale=0.98]{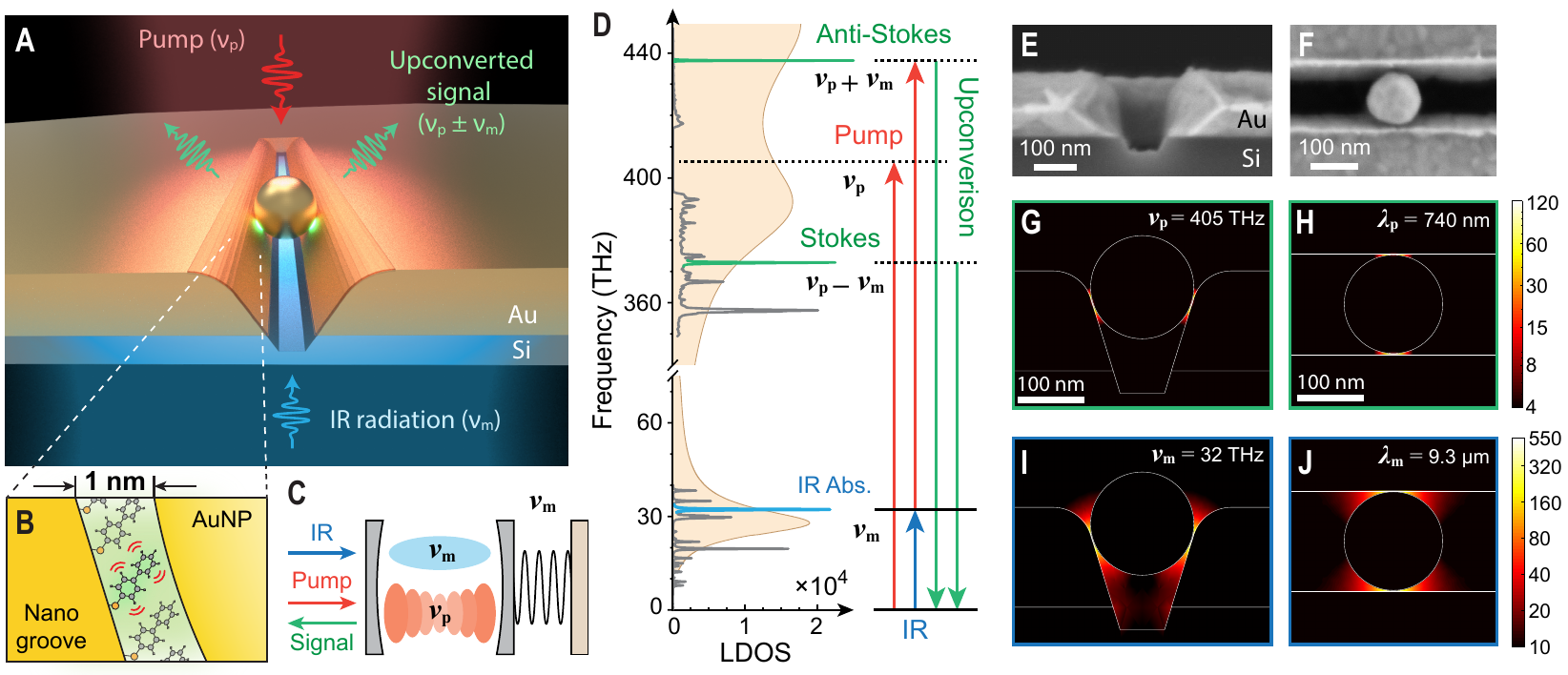}
    \caption{
      \textbf{Molecular optomechanical upconversion concept}.
    (\textbf{A}) Illustration of a nanoparticle-in-groove molecular optomechanical cavity which confines both IR (frequency $\nu_\mathrm{m}$) and visible (frequency $\nu_\mathrm{p}$) fields into the $\sim$1-nm-thick molecular layer (\textbf{B}). (\textbf{C}) Molecular vibrations couple resonantly to the IR field because they change the molecular dipole moment, while the change in the molecule's polarizability leads to a parametric coupling to the visible field, realising an optomechanical cavity.
    (\textbf{D}) Resonant vibrational levels and electromagnetic fields involved in the upconversion together with the computed radiative local density of state (LDOS) inside the nanocavity, along with the measured Raman scattering and simulated IR absorption (cf. Sec.~\ref{sec:molecules}) of the molecules. 
    (\textbf{E}) Cross sectional view of scanning electron microscope (SEM) images of a fabricated nanogroove.
    (\textbf{F}) Top view SEM image of a nanoparticle-in-nanogroove.
    (\textbf{G-J}) Simulated electromagnetic field enhancement factor for incident plane waves at $\nu_\mathrm{p}=405$~THz (740~nm, \textbf{G,H}) and $\nu_\mathrm{m}=32$~THz (9.3~$\mu$m, \textbf{I,J}), both polarized orthogonal to the groove.
     }
    \label{fig:concept}
\end{figure}

\paragraph*{Device description}

The idea of a molecular optomechanical platform for upconversion was first proposed in \cite{roelli2016} and its theoretical performance was analysed in \cite{roelli2020}, showing the feasibility of single-photon sensitivity at frequencies down to few THz even at ambient conditions. 
In contrast to plasmomechanical resonators that have been realised e.g. with dimer antennas built on a nanobeam oscillator \cite{thijssen2015} or with nanoparticles trapped in a plasmonic hotspot \cite{mestres2016}, the mechanical resonator in a molecular optomechanical cavity  consists in a collective molecular vibration, which is parametrically coupled to the nanocavity through its Raman polarizability \cite{roelli2016}. 
This approach allows reaching mechanical frequencies in the 1--100~THz range and optomechanical coupling rates above 1~THz \cite{benz2016a}. 

Our experiment is conceptually presented in Fig.~\ref{fig:concept}A-D.
To obtain extreme field confinement in nm-scale regions as well as large enhancement factors at all frequencies involved in the upconversion, we develop a new type of plasmonic nanocavity that we dub the nanoparticle-in-groove (Fig.~\ref{fig:concept}E-F).
It consists of a single Au nanoparticle (150~nm nominal diameter) placed inside a nanogroove etched in a Au film and covered by a monolayer of biphenyl-4-thiol (BPhT) acting as spacer and molecular oscillator (Fig.~\ref{fig:concept}B-C).
Our structure is engineered to support colocalized plasmonic resonances at IR ($\sim$30~THz) and visible frequencies, which can be excited under normal incidence illumination (more details are given in SM Sec.~\ref{sec:EMsimulation} and Fig.~\ref{fig:simulation}). 
The IR resonance frequency is governed by the length of the nanogroove \cite{garcia-vidal2010,huck2015}, which was chosen as 2~$\mu$m to match a vibrational mode of BPhT molecules at 32 THz with large IR and Raman cross sections (see SM Sec.~\ref{sec:molecules}).
Two broad resonances in the visible domain are supported by the nanoparticle coupled to the walls of the Au nanogroove, analogous to the nanoparticle-on-mirror geometry \cite{ciraci2012,baumberg2019}. 
Near fields are confined in the nanometer-wide gaps formed by the molecular layer (see Fig.~\ref{fig:concept}G-J).
This design provides enhancement factors for IR absorption and Raman scattering up to $\times 10^5$ and $10^8$, respectively (estimated from the second and fourth power of the maximal field enhancements).
Overall, our geometry is expected to boost the conversion efficiency by a factor on the order of $10^{13}$ for the molecules located in the region of highest field, compared to free space.

\paragraph*{Results}
\begin{figure}
    \centering
    \includegraphics[scale=1]{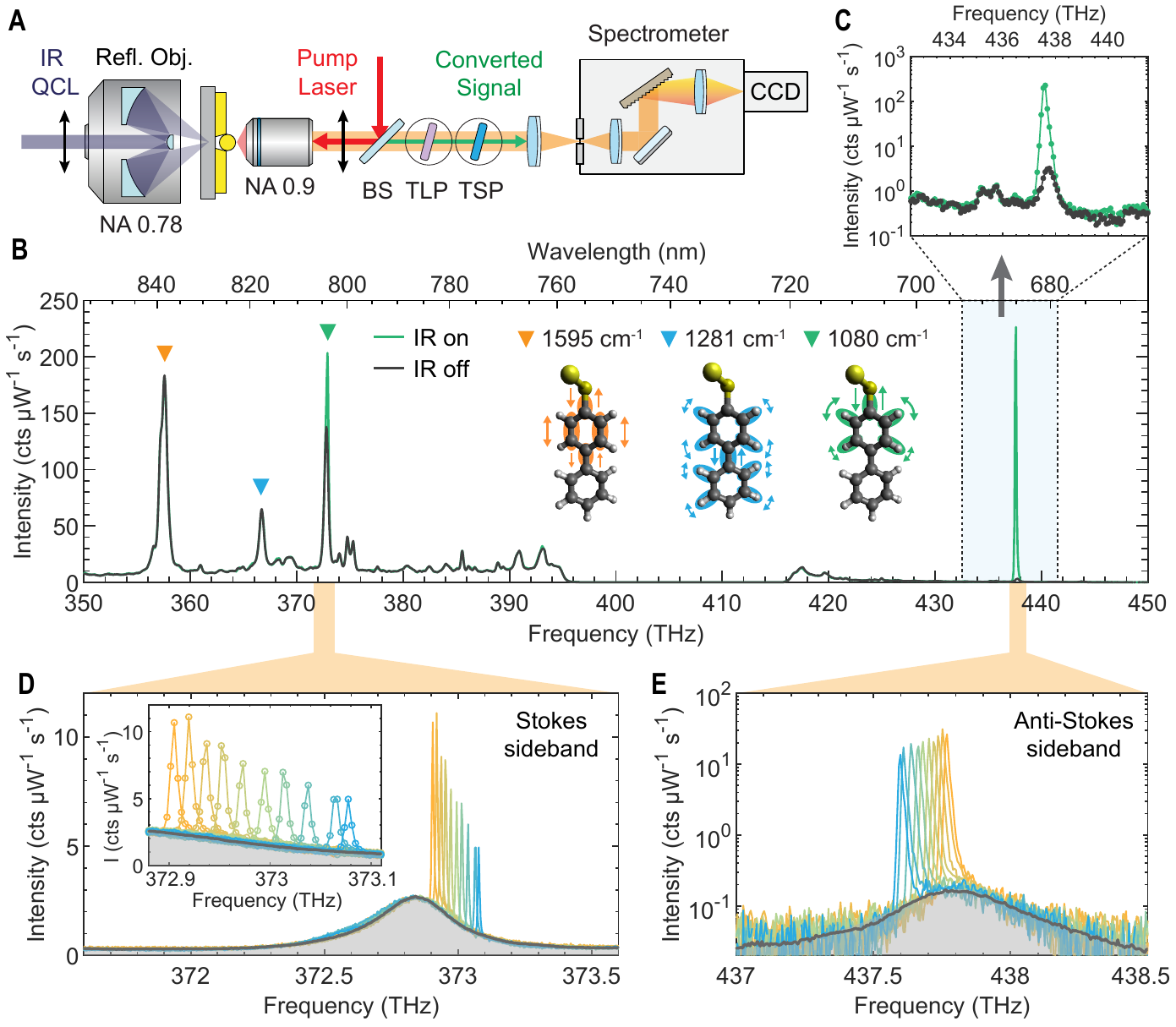}
    \caption{
\textbf{Molecular optomechanical transduction from 32~THz to the visible domain.} (\textbf{A}) Schematic of the measurement configuration. A reflective objective with numerical aperture (NA) 0.78 focuses the IR beam through the silicon substrate, while a refractive objective (NA 0.9) focuses the visible pump beam and collects the Raman signal, which is directed to a spectrometer after blocking the pump light with tunable short- or long-pass filters (TSP, TLP). The polarization of both pump and IR radiation is perpendicular to the nanogroove (black arrows) for effective excitation of the plasmonic resonances. BS: beam splitter.
    (\textbf{B}) Low-resolution, full-range Raman spectra from a single nanocavity under 10~$\mu$W pump power (740~nm wavelength) without (black line) and with (green line) incoming IR radiation  (530~$\mu$W) resonant with the vibrational mode at 1080~cm$^{-1}$ or about 32~THz. The inset shows the displacements of typical Raman-active modes labelled with orange, blue and green triangles. Acquisition time, 10~s. 
    (\textbf{C}) Closer view on the anti-Stokes sideband of panel \textbf{B} plotted on a logarithmic vertical scale.
    (\textbf{D,E}) High resolution (0.23~cm$^{-1}$, 7~GHz) Stokes (\textbf{D}) and anti-Stokes (\textbf{E}) spectra observed when tuning the signal beam (range limited by the QCL capabilities), all normalized to 175~$\mu$W incoming IR power (pump: 10~$\mu$W). 
    The grey line in \textbf{D} and \textbf{E} are the spectra without IR radiation, where the grey areas are the corresponding Lorentzian fits to the spontaneous Stokes and anti-Stokes scattering peaks. The inset of \textbf{D} shows the enlarged view of the resolution-limited lineshape of the upconverted signal (see also Figs.~\ref{fig:ProfileFit},\ref{fig:Fit_VS}).  Acquisition time for the upconverted signal (on both sides), spontaneous Stokes and anti-Stokes spectra are 10~s, 300~s and 600~s, respectively.
}
    \label{fig:spectrum}
\end{figure}

When a pump laser tuned at 405~THz (740~nm) is focused on the sample (Fig.~\ref{fig:spectrum}A), the parametric interaction with the molecular vibrations generates Raman sidebands at lower (Stokes) and higher (anti-Stokes) frequencies (Fig.~\ref{fig:spectrum}B). 
The giant enhancement factor described above makes it possible to detect the Raman signal from few hundreds molecules \cite{langer2020}, as estimated from the mode volume (Fig.~\ref{fig:simulation}) and molecular layer density \cite{ahmed2021} (see also Sec.~\ref{sec:theory}).
Without IR beam incident on the device (black solid line in Fig.~\ref{fig:spectrum}B), the Stokes signal is dominated by spontaneous emission of phonon-photon pairs while the anti-Stokes signal originates from the upconversion of thermal vibrations \cite{velez2019}. 
At room temperature ($T=25^\circ$C), the thermal occupancy of the vibrational mode at $\nu_\mathrm{m}=32$~THz is $n_\mathrm{th}= \left(\exp\left(\frac{h\nu_\mathrm{m}}{k_\mathrm{B}T}\right)-1\right)^{-1} \simeq 5.8\times 10^{-3}$, where $h$ and $k_\mathrm{B}$ are Planck's and Boltzman's constants.

When the IR beam from a quantum cascade laser (QCL) is focused through the Si substrate onto the backside of the device  (Fig.~\ref{fig:spectrum}A), we observe the amplification of a single resonant peak in the Stokes and anti-Stokes sidebands (green line in Fig.~\ref{fig:spectrum}B), whose linewidth is much narrower than the natural linewidth of spontaneous Raman scattering peak (Fig.~\ref{fig:spectrum}D,E).
Altogether, these observations are compatible with a coherent upconversion process.
As we tune the frequency of the incoming field the upconverted signal shifts accordingly, and its measured linewidth is found to be limited by that of our spectrometer, with a value of 7~GHz or 0.23~cm$^{-1}$ (cf. Fig.~\ref{fig:Fit_VS}), which is well below that of single-molecule Raman linewidths typically observed at room temperature \cite{artur2011} (see Sec.~\ref{sec:linewidth} for further discussion).
The relative conversion efficiency vs. detuning is plotted in Fig.~\ref{fig:ProfileFit} and confirms that upconversion is assisted by the vibrational mode. 

We interpret these results as the manifestation of optomechanical transduction, where a collective molecular vibrational mode is resonantly driven by the nanocavity-enhanced 32~THz incoming field. 
We can describe the resulting vibrational state by a displaced thermal state with a coherent amplitude $\alpha$ and a corresponding mean phonon number $n_\mathrm{coh}=|\alpha|^2$.
This oscillation, which is coherent with the IR drive, is mapped onto the Raman sidebands of the pump laser, where the IR signal can be analysed and detected using a standard optical spectrometer, camera or single photon counting module.
A lower bound for $n_\mathrm{coh}$ is estimated from the ratio of the Raman peak areas with vs. without IR drive. 
It is a lower bound as not all molecules that contribute to the spontaneous Raman peak do also contribute to upconversion, due to the imperfect overlap of IR and VIS plasmonic modes. 
For most nanocavities, we find IR-driven phonon numbers $n_\mathrm{coh}>0.1$ for 500 to 600 $\mu$W IR power, equivalent to $\frac{n_\mathrm{coh}}{n_\mathrm{th}}>10$ (cf. Table~\ref{tab:statistics}). 
This figure is compatible with an enhancement of IR absorption cross-section per molecule by more than 5 orders of magnitude, as simulated in Fig.~\ref{fig:concept}I,J and detailed in Sec.~\ref{sec:theory} in SM, corresponding to IR photon-to-vibration conversion efficiency above $10^{-5}$.
 
\begin{figure}
    \centering
    \includegraphics[scale=1.2]{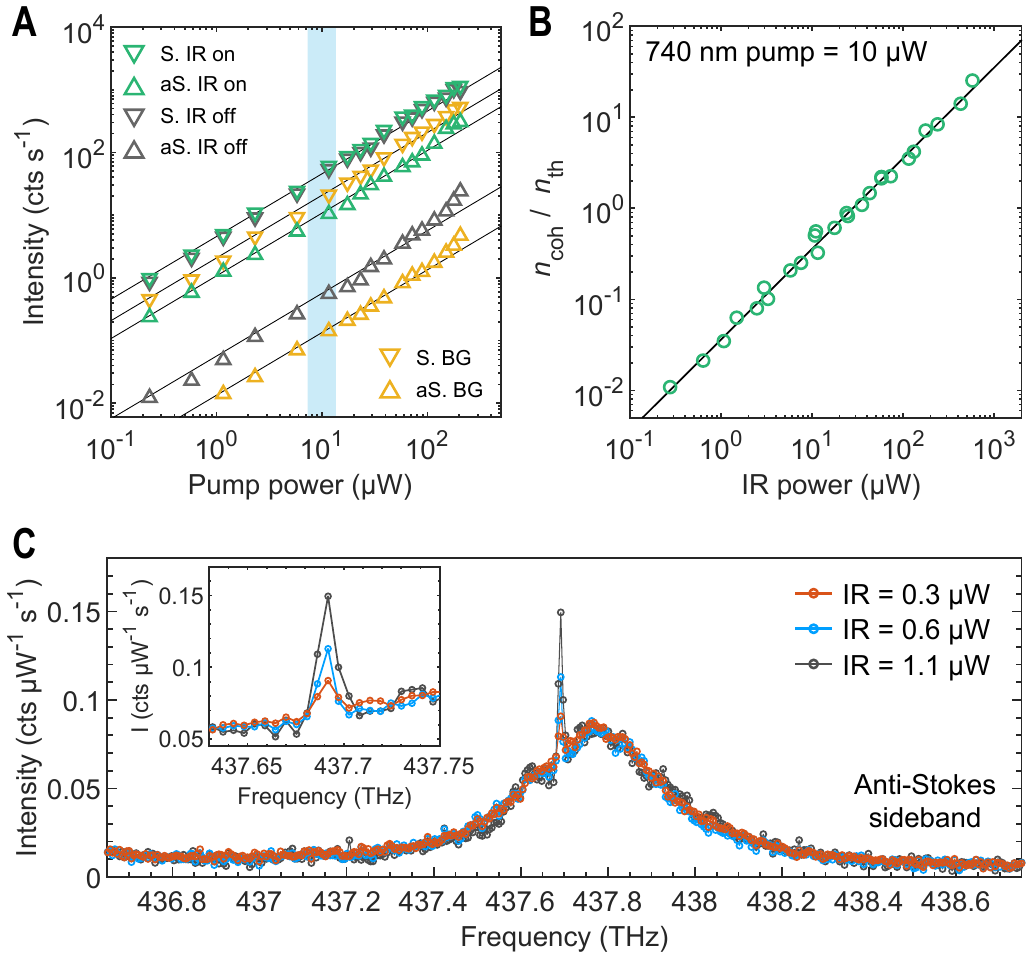}
    \caption{
    \textbf{Dependence of upconverted signal on mid-IR and pump powers.}
    (\textbf{A}) Measured Raman signal as a function of pump power (tuned at 740~nm). Triangles pointing up (resp. down) correspond to anti-Stokes (resp. Stokes) sideband. Green symbols correspond to an incident IR power of 600~$\mu$W, while grey symbols are taken without IR drive. The spectrally broad background underlying the Raman peaks is shown with orange symbols. 
    Black lines are linear fits for reference. The blue shaded area denotes the power range used in typical upconversion measurements presented here.
    (\textbf{B}) Ratio of IR-driven ($n_\mathrm{coh}$) to thermal ($n_\mathrm{th}$) phonon occupancies as a function of the IR radiation power on the nanocavity (measured with 10~$\mu$W pump power, tuned at 740~nm). The black line shows a linear fit.
    (\textbf{C}) Detection of sub-$\mu$W IR signals using high-resolution spectroscopy on the anti-Stokes sideband. Acquisition time, 600~s.
   }
    \label{fig:power}
\end{figure}

Next, we characterise the power dependence of the upconverted signal, by varying first the power of the pump laser at 740 nm (Fig.~\ref{fig:power}A). 
We observe a linear dependence of the upconverted Raman signal, in adequation with the expected parametric upconversion process.
Moreover, in the measurement conditions of Fig.~\ref{fig:spectrum}, the spontaneous (no IR drive) Stokes and anti-Stokes signals also grow linearly with pump power.
An optomechanical description of plasmon-enhanced Raman scattering \cite{roelli2016,schmidt2016b,roelli2020} predicts three main regimes for the pump power dependence of the anti-Stokes signal: (i) at low power a linear regime dominated by thermal noise, then (ii) a quadratic regime where the vibrational population increases linearly with laser power due to quantum backaction (also called vibrational pumping) \cite{kneipp1996,benz2016a,schmidt2017}, and finally (iii) a phonon-stimulated regime is expected,  dominated by dynamical backaction amplification of the vibration \cite{roelli2016}.
The data in Fig.~\ref{fig:power}A are compatible with regime (i) above, i.e. we do not observe a superlinear dependence of the anti-Stokes signal on laser power in the range used in the upconversion experiment (blue shaded area).
Fig.~\ref{fig:power}B shows how the upconverted signal scales with incoming mid-IR power, with results compatible with a linear dependence, as expected for a resonant drive below saturation. 

Finally, we quantify the external IR to visible conversion efficiency of our device by spectrally filtering the anti-Stokes sideband and sending it to a single photon counting module, with independently calibrated detection efficiency (results are summarised in Table.~\ref{tab:statistics}).
Taking into account the calibrated detection efficiency, we infer that the upconverted anti-Stokes photon rate collected by the objective reaches up to 200~kHz per nanocavity for 600~$\mu$W incident IR power (corresponding to $n_\mathrm{IR}\simeq 2.8 \times 10^{16}$ photons per seconds) and 10~$\mu$W pump power.
This measurement yields a conversion efficiency from incoming IR photon to outgoing visible photon collected by our objective on the order of $10^{-12}$ (see Fig~\ref{fig:eta_histogram}), or $10^{-10}$~mW$^{-1}$ as normalised per unit pump power.
We note that the only 6\% of the power emitted in the near field is collected by our objective, as detailed in Fig.~\ref{fig:simulation}E, so that the internal efficiency is at least 10 times larger. 
Despite the low efficiency, the coherent nature of the process allows us to reliably detect incoming IR powers down to few hundreds nanowatts, as shown in Fig.~\ref{fig:power}C -- a figure which would further improve with the resolution of the spectrometer.

\paragraph*{Discussion}
To conclude this Report, we discuss a few avenues for future research and development in molecular optomechanical frequency conversion. 
A possible route for efficiency improvement consists of increasing the number of molecules contributing to optomechanical interaction without sacrificing on field enhancement, which could be achieved by increasing the number of nanoparticles in a singe nanogroove, optimising the shape and facet size of the nanoparticles, increasing the density of grooves and possibly leveraging lattice resonances.
In this spirit, new dual-resonant metamaterials \cite{wang2007,le2008,mueller2021} with nanocavities as unit cells could be developed to enhance IR absorption \cite{yao2014} and upconversion efficiency.
We note that even with a single nanocavity, internal conversion efficiencies approaching 0.1 were predicted in Ref.~\cite{roelli2020} by operating at higher pump powers under pulsed excitation \cite{lombardi2018}, allowing for single THz/mid-IR photon detection under pulsed pumping.

We also predict that leveraging strong light-matter coupling to boost the IR absorption and/or Raman scattering will become a particularly fruitful area of research for upconversion applications. 
In the vibrational strong coupling regime and under critical cavity input coupling, most absorbed radiation can be transfered to vibrational polaritons \cite{shalabney2015}, so that even single-photon-level signals could be efficiently upconverted \cite{roelli2020}.
Conversely, there has been recent indications that the formation of exciton-polaritons in nanocavities similar to the one used here can further enhance the resonant Raman cross-section \cite{liu2021}, offering a way to improve the internal upconversion efficiency.
In addition to plasmonic resonance engineering, molecular engineering \cite{shao2006} also offers promising perspectives for improving IR and Raman cross-section, controlling the conversion bandwidth, and addressing a broader range of frequencies down to the THz domain. 
Optical phonons in 2D materials are other candidates for reaching lower frequencies,
while two-photon driving of a vibration could be explored to extend operation to frequencies lower than that of the vibrational resonance \cite{maehrlein2017,traverso2018}.

For a better fundamental understanding of degree of coherence and added noise in the transduction process, measurements of first- and second-order correlation functions as in e.g. \cite{forsch2020} can be performed.
Besides, broader tuning of the IR and pump tones and higher resolution spectroscopy should reveal new information about the molecule-nanocavity interaction and clarify possible non-resonant contributions to upconversion related to the electronic response of metal and molecules \cite{shen1989,tian2014}.
Looking ahead, the sub-wavelength size of our device will allow for the construction of chip-scale pixel arrays, spectrometers and hyperspectral imagers for THz and IR wavelengths, with the  possibility to excite and detect the upconverted signals through dielectric waveguides \cite{losada2019}.
And since the transduction mechanism is bidirectional, it can be used for coherent downconversion as well, allowing optical generation of long-wave radiation assisted by molecular systems \cite{hasselbeck2004,delpino2016}.

\bibliography{midIRup_zot}
\bibliographystyle{ieeetr}
\section*{Acknowledgments}
C.G. is indebted to Vivishek Sudhir  for valuable comments and fruitful discussions about the results.  
W.C. and C.G. acknowledge support from IPHYS mechanical workshop characterization platform as well as CMi cleanroom at EPFL. 

\paragraph*{Funding}
This work received funding from the European Union’s Horizon 2020 research and innovation program under Grant Agreement No. 829067 (FET Open THOR) and Grant Agreement No. 820196 (ERC CoG QTONE).
C.G. acknowledges the support from the Swiss National Science Foundation (project numbers 170684 and 198898).
It is part of the research program of the Netherlands Organisation for Scientific Research (NWO).

\paragraph*{Author contributions}
W.C. designed and fabricated the devices, performed the experiments, analysed the data and created the main figures.
P.R. performed calculations of molecular parameters, assisted in early stages of the experiments and in data analysis.
H.H. performed the electromagnetic simulations of nanocavities and contributed to nanocavity design.
S.V. assisted in recording and analysing photon counting data. 
S.P.A. contributed to setting up the experimental apparatus.
A.I.B., M.K. and A.M. conceived and fabricated the first generation of nanogroove cavities.
T.J.K. contributed to early ideas leading to this work and commented the manuscript.
E.V. and A.M. discussed the results and contributed to write and improve the manuscript.  
C.G. designed and supervised the study, analysed the data and wrote the manuscript with the assistance of W.C., P.R., H.H., E.V. and A.M.

\paragraph*{Competing interests}
The authors declare no competing interests.

\paragraph*{Data and materials availability}
All data supporting this Report will be made available on a Zenodo repository.  

\section*{Supplementary materials}
Materials and Methods\\
Supplementary Text\\
Figs. S1 to S10\\
Tables S1 to S2\\
References \textit{(67-74)}

\newpage
\setcounter{figure}{0}  
\renewcommand{\figurename}{\textbf{Supplementary Figure}}
\renewcommand{\tablename}{\textbf{Supplementary Table}}
\renewcommand{\thefigure}{S\arabic{figure}}
\renewcommand{\thetable}{S\arabic{table}}

{\huge Supplementary materials}

\section{Materials and Methods}

\subsection{Nanocavity fabrication}\label{sec:fabrication} 

The device where fabricated using a double polished silicon wafer (380-nm-thick) as the substrate. A 2-nm-thick Cr adhesion layer and a 150-nm-thick gold layer were thermally evaporated on the Si at a rate of 0.5~nm/s. An array of 150~nm wide and 2~$\mu$m long nanogrooves with unit cell 4$\times$5~$\mu$m$^2$ was patterned by electron-beam lithography using poly methyl methacrylate (PMMA) as photoresist. After photoresist development, the nanogrooves were etched by angled collimated ion-beam, resulting in the cross-section shown in Fig.~\ref{fig:concept}E. The sample was then immersed in piranha solution (H$_2$SO$_4$: H$_2$O$_2$ = 3:1) for 5 min to remove PMMA, which also annealed the Au to smooth the surface.  The wafer with nanogrooves was subsequently diced into chips. The single chips were cleaned with ozone for 30~min under ultraviolet light illumination and then put into ethanol solution overnight to remove Au oxide. The sample was then incubated into a 1~mMol solution of Biphenylthiol (BPhT, Sigma-Aldrich) in ethanol for 24~h at ambient conditions to form a self-assembled monolayer \cite{ahmed2021}, followed by 5 rounds of ethanol rinsing and drying by nitrogen gas flow.
Finally, a drop of aqueous solution containing 150~nm diameter gold nanoparticle (BBI solution)  was deposited on the nanogroove region. After 10 min incubation, the liquid was removed by nitrogen gas flow and the sample was then rinsed by de-ionized water and dried by the gas flow. The nanoparticles were randomly distributed on the Au film and inside the grooves. 
We found that nanoparticles at any relative position inside a groove enables frequency upconversion (see the inset of Fig.~\ref{fig:TSspectra}A, and also the statistics in Table~\ref{tab:statistics}).

\subsection{Optical setup and data acquisition}\label{sec:acquisition} 

\begin{figure}
    \centering
    \includegraphics[scale=1.1]{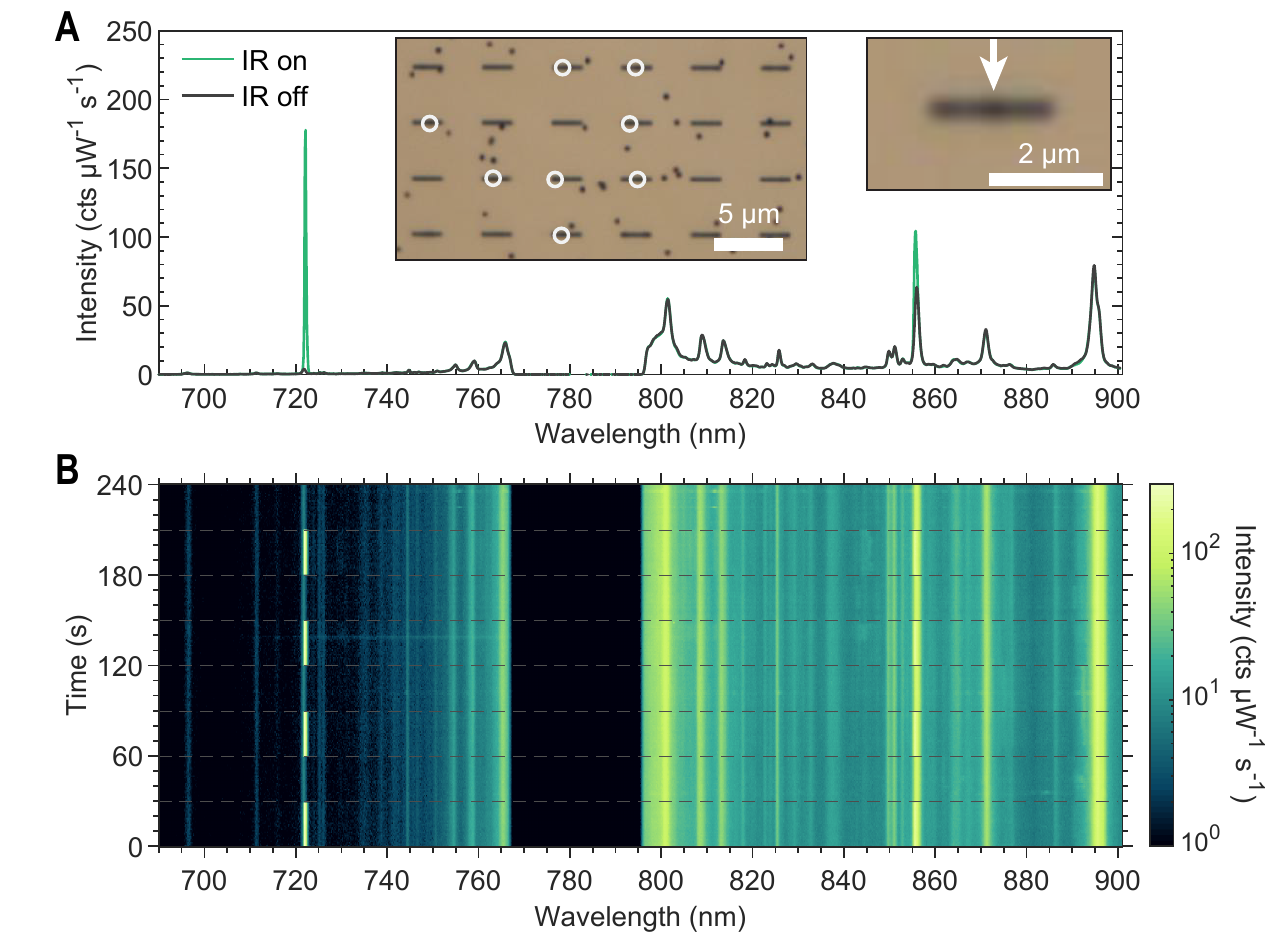}
    \caption{\textbf{Optical image of the nanoparticle-in-nanogroove system and time series of the Raman spectra with and without IR drive.}
    (\textbf{A,B}) Raman spectra of a nanocavity with and without IR drive (\textbf{A}) under pumping at 785~nm , which were obtained from the time series of the Raman spectra (\textbf{B}) to show the stability and reproducibility of the upconversion measurement. The inset in \textbf{A} shows a typical region of the sample, where the single 150~nm-diameter nanoparticle inside the nanogroove (labelled by white circles) can be identified due to the strong scattering cross section of the nanoparticle. Note that no upconverted signal can be observed from the nearby nanoparticles sitting on the Au film. Pump and IR powers are 8.6~$\mu$W and 610~$\mu$W, respectively. Acquisition time, 10~s.
}
    \label{fig:TSspectra}
\end{figure}

\begin{figure}
    \centering
    \includegraphics[scale=1.4]{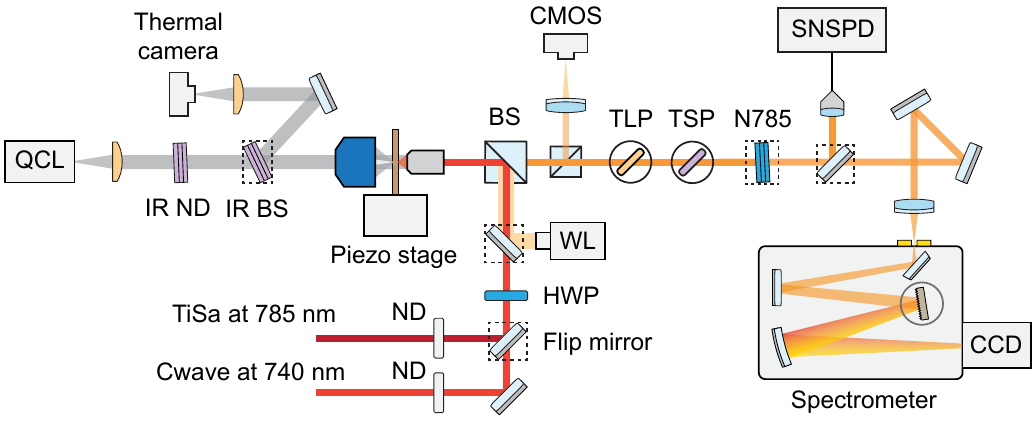}
    \caption{\textbf{Schematic of the optical setup.}
    QCL: quantum cascade laser; ND: neutral density filter; BS: beam splitter; CMOS: complementary metal–oxide–semiconductor camera (mvBlueFOX); TLP: tunable long-pass filter; TSP: tunable short-pass filter; N785: notch filter at 785 nm; SNSPD: superconducting nanowire single-photon detector; WL: white light; HWP: half-wave plate; CCD: charge-coupled device.
    }
    \label{fig:setup}
\end{figure}

We built a custom transmission microscope as schematised in Fig.~\ref{fig:spectrum}A and detailed in Fig.~\ref{fig:setup}. The sample was at ambient temperature in air, mounted on closed-loop XYZ piezo-positioners (AttoCube). As shown in the inset of Fig.~\ref{fig:TSspectra}A, single nanoparticles inside nanogrooves can be observed directly under bright field white light illumination.

For the excitation of Raman scattering, the pump laser was focused by a refractive objective (Olympus, max. numerical aperture NA 0.9) from the side of the gold film.
We used two types of lasers as the pump light: (i) A continuous-wave Ti:Sapph laser (Model 3900 SpectraPhysics) tuned around 785~nm, where we can use a notch filter at 785 nm (Thorlabs) in detection to detect both Raman sidebands simultaneously;
(ii) a C-WAVE OPO from Hubner Photonics tuned around 740~nm, where the spectra of Stokes and anti-Stokes sidebands were taken sequentially by introducing tunable long-pass and short-pass filters (Semrock), respectively. 
While the Ti:Sapph laser features a measured linewidth of $20$~GHz or more with spectral jitter leading to a Gaussian lineshape, the C-WAVE has a specified linewidth of 1~MHz. 
Using the C-WAVE line we find that the resolution of our spectrometer is at best 7~GHz, defined as the measured FWHM (Fig.~\ref{fig:Fit_VS}I), so that it is the limiting factor in the measurements of Fig.~\ref{fig:spectrum}D-E.

Using a broadband half-wave plate, the polarization of the pump light was tuned perpendicular to the groove's long-axis to effectively excite the plasmonic modes in the visible domain (Fig.~\ref{fig:spectrum}A). Without additional specification, the power of both lasers for all the measurement were kept at the level of $\sim$10~$\mu$W on the sample to avoid slow nanocavity modifications and other nonlinearities that we observe above $\sim$10~$\mu$W (see Fig.~\ref{fig:power}A).
Reflected and scattered fields were collected through the same refractive objective, and then spectrally filtered to isolate Stokes and anti-Stokes Raman sidebands, using a notch or two different edge filters (see above), before being focused on the entrance slit of a spectrometer, or alternatively coupled to a telecom fiber and routed to super-conducting nanowire single photon counting modules held in a closed-cycle cryostat below 2~K. 
We use two spectrometers with 3 different configurations: a Kymera 193i (grating 600 line/mm) and Shamrock 750 (grating 150 line/mm) for broad spectral range and lower resolution (as in Fig.~\ref{fig:spectrum}B-C and Fig.~\ref{fig:NPstatistics}) and the Shamrock 750 with 1800 line/mm grating for higher resolution (used in Fig.~\ref{fig:spectrum}D,E and Fig.~\ref{fig:TiSa785_QCLTune}B,C). 

A single mode quantum cascade laser (QCL, Alpes Laser) provides the coherent IR signal for the upconversion measurements. Its specified tuning range is from $\sim$1071 to $\sim$1077~cm$^{-1}$ (see detailed performance measured by upconversion in Fig.~\ref{fig:QCLpeformance}), close to the center of the Raman peak used in upconversion at $\sim$ 1080~cm$^{-1}$ (we recall that 1~cm$^{-1}\simeq 30$~GHz). 
The QCL is collimated by a ZnSe lens and then focused through the silicon substrate (approximately 30\% transmission) on a nanocavity with a reflective objective (PIKE technologies, max. NA 0.78). To excite the IR cavity mode, the orientation of the nanogrooves is set perpendicular to the QCL polarisation.
The alignment of the IR beam can be adjusted by observing the IR light reflected from the Au film on Si, where a part of the IR light was directed to a thermal camera (Seek Thermal) using an IR beam splitter (Thorlabs), allowing to check the beam quality (cf. Fig.~\ref{fig:setup}). The IR beam size on the sample is $\sim$10~$\mu$m. 

Control measurements were also performed on (i) the nearby nanoparticle-on-mirror system outside a groove (inset of Fig.~\ref{fig:TSspectra}) and (ii) the bare nanogroove without nanoparticle; but also on a nanoparticle-in-nanogroove with either (iii) IR radiation or (iv) pump light polarised parallel to the long-axis of the groove. In all these cases, no upconverted signal could be observed, confirming the essential role of the overlapped plasmonic visible and IR resonances for efficient optomechanical transduction. 

\paragraph{IR power dependence}
The QCL power can be tuned from $\sim$3.6~mW to $\sim$60~$\mu$W (the lower limit of the bolometer) before the reflective objective by cooperatively varying its working temperature and drive current (see Fig.~\ref{fig:QCLpeformance}). Given that the transmission through the objective and the Si substrate in our system are 55\% and 30\%, respectively, the actual IR power on the nanocavity can be adjusted in this way between $\sim$600~$\mu$W and $\sim$10~$\mu$W. Combing this tunability with a series of IR neutral density filters (Thorlabs), the lowest power in our measurement was $\sim$0.3~$\mu$W (Fig.~\ref{fig:power}C). The IR power dependent spectra were taken using high resolution spectrometer on the anti-Stokes band excited by 740~nm pump light. We used acquisition times ranging from 10 to 600~s, with the upper limit set by the mechanical drift of the stage holding the sample.

\paragraph{IR frequency scanning}
For the measurements in high-resolution mode shown in Fig.~\ref{fig:spectrum}D,E and ~\ref{fig:TiSa785_QCLTune}B,C, the IR radiation was tuned from $\sim$1071 to $\sim$1077 cm$^{-1}$ (32.1 to 32.3~THz) while the IR power is in the range from $\sim$80 to $\sim$200 $\mu$W on the nanocavity. The acquisition time for the Raman signal with IR radiation is 10~s. The acquisition time for Anti-stokes and Stokes signal without IR radiation are 600~s and 300~s, respectively. The fluctuation of the raw spectra were offset based on keeping the spontaneous peak the same height. 
The differences in IR powers as a function of frequency was corrected by the following treatment: (i) the narrow peak from the net upconverted signal and the broad peak from the spontaneous Raman were separated mathematically; (ii) the narrow peak intensities for different QCL frequencies were normalised to $\sim$175 $\mu$W by linearly scaling the spectra (given that the IR power dependence is linear, see Fig.~\ref{fig:power}B); (iii) the normalised narrow peaks were added back to their spontaneous background.
From this measurement, we can also plot the relative upconversion efficiency vs. IR frequency, with results following the  lineshape of spontaneous Raman scattering (Fig.~\ref{fig:ProfileFit})



\subsection{Measurement of upconversion efficiency} \label{sec:calibration} 
To estimate the end-to-end upconversion efficiency, we must measure as precisely as possible the upconverted photon flux, taking into account all detection losses in the Raman signal path.
For this purpose, we replace the interference filters (TLP, TSP and N785 in Fig.~\ref{fig:setup}) in the detection path by neutral density filters with known optical densities, measure the laser power reflected from the sample, from which we compute the collected photon flux, and compare this photon flux to the peak-integrated count rate on the spectrometer's CCD.
From this calibration method, we can relate the upconverted anti-Stokes Raman spectrum to the corresponding anti-Stokes photon flux collected by the objective. 
We find that 1~Hz count rate on the CCD (with the low-resolution grating) corresponds to $\sim$50~Hz collected photon rate.

In typical experimental conditions ($\sim$ 10~$\mu$W pump power on sample), after correcting for the measured detection efficiency, the thermal anti-Stokes photon rate collected by the objective is found to be about 1 to 4~kHz depending on nanocavity,
while the photon rate reaches more than 100~kHz for 600~$\mu$W IR power for the most efficient devices (see Table~\ref{tab:statistics}).

To further confirm the values extracted above, we use fiber-coupled superconducting nanowire single-photon detectors (SNSPD, Single Quantum) connected to a time-to-digital converter (ID900, ID Quantique) for count rate readout.
To minimize the spontaneous Raman background we measure the anti-Stokes sideband, proceeding as follows. The Raman peak resonant with IR drive is filtered by a combination of tunable short and long-pass filters from the nearby peaks. The filtered beam is coupled into a telecom fiber connected to the SNSPDs.
By accounting for the transmission of objective and beam splitter (62.5\%) and of the filters tuned to their edges ($<30$\%), for the fiber coupling efficiency (at best 40\%) and the SNSPD efficiency at this wavelength (less than 50\%), we expect an overall conversion factor from collected photon rate to count rate on the order of 3 to 4\%. 
The measured count rates in this way are on the order of 1 to 5~kHz, which is therefore in good agreement with the estimated anti-Stokes photon fluxes from the method described above in Table~\ref{tab:statistics}.

To conclude this section, we mention that from the simulations described in Sec.~\ref{sec:EMsimulation} only 6\% of the power upconverted inside the nanocavity is actually collected by our objective (the remainder is either radiated in other directions or dissipated in the metal, see Fig.~\ref{fig:simulation}).
All measured efficiency do not account for this factor, and are therefore conservative estimates. 

\subsection{Electromagnetic simulations} \label{sec:EMsimulation} 

Electromagnetic simulations were performed with commercial FEM package COMSOL Multiphysics 5.2. The 3D structure contained a 155~nm gold nanoparticle situated on a 2 $\mu$m nanogroove whose trapezoid cross-section had a 180~nm top-base and a 65~nm bottom-base. The height of the nanogroove cross-section is 185~nm, including a 35~nm dip in the silicon substrate (determined from SEM images). The nanogroove was carved into a 150~nm thick gold layer which was situated on an infinitely thick silicon substrate. The dielectric function of gold follows the data from Johnson and Christy (VIS/NIR domain,\cite{johnson1972}) and Olmon et al. (IR domain,\cite{olmon2012}). The refractive index of the silicon is set as 3.4 in the IR range, and follows Green et al. in the visible range \cite{green2008}. 

To calculate the radiative emission and infer the local density of states (LDOS, Fig.\ref{fig:simulation}A, C), an electric dipole $\mathbf{\mu}_0$ was placed inside the nanogap formed by the molecular layer. The radiated power $P_{\mathrm{rad}}$ collected within a given solid angle determined by the numerical aperture of the objective (visible, NA = 0.9; IR, NA = 0.78) was normalized by the dipole radiation in vacuum $P_0$ as $P_{\mathrm{rad}}/P_0$. 
The radiative emission from the dipole in the nanogap $P_{\mathrm{rad}}$ was calculated as the surface integral of the Poynting vector in the far field,\cite{yang2016} while the dipole radiation power in vacuum reads $P_0=\frac{|\mathbf{\mu}_0|^{2}}{4 \pi \varepsilon_{0} \varepsilon} \frac{n^{3} \omega^{4}}{3 c^{3}}$. 
The ratio of power radiated within the objective's NA to the total vacuum emission is identified as the radiative LDOS enhancement multiplied by collection efficiency $\eta_{\mathrm{coll}}$, $\rho_{\mathrm{rad}}/\rho_0=P_{\mathrm{rad}}/P_0 \cdot\eta_{\mathrm{coll}}$.
Radiative emission towards the silicon substrate was calculated similarly. 

The nonradiative decay is attributed to Ohmic loss (absorption), which was calculated by the total decay subtracted by the radiative decay. Here, the total decay rate was calculated by a closed surface integral only containing the electric dipole in the gap. The antenna radiative efficiency can be obtained as $\eta_\mathrm{rad}=P_{\mathrm{rad}}/P_{\mathrm{tot}}$, where $P_{\mathrm{tot}}$ is the above-mentioned total dipole radiation in the nanogap. 

\subsection{Molecular calculations}\label{sec:molecules} 

Raman and infrared cross-sections are calculated with density functional theory (DFT) using \textsc{Gaussian09} (B3LYP/6-311++G(d,p) base). The choice of molecular geometry is based on the comparison of the calculated DFT vibrational spectrum for BPhT alone and BPhT bound to gold (Au-BPhT) with the experimental results (Fig.~\ref{fig:dft}A), which shows better agreement with Au-BPhT.  
The orientation of the Au-BPhT molecule with respect to the local fields and the linewidth of the vibrational modes are similarly adapted to best match experimental data. 
The vibrational modes' IR absorption cross-section ($\sigma_{IR}$, Fig.~\ref{fig:dft}B) and differential Raman cross-section ($d\sigma_R/d\Omega$, Fig.~\ref{fig:dft}C) are calculated from the derivatives of the polarizability tensor and electric dipole with respect to the displacements of the molecule's atoms, following the procedures described in Refs.~\cite{ru2009,roelli2020}.

\subsection{Theoretical predictions of conversion efficiency from calculated IR and Raman cross-sections} \label{sec:theory} 

In this section we predict the upconverted Raman power $P_\mathrm{up}$ (photon flux $\phi_\mathrm{up}=P_\mathrm{up}/h\nu_\mathrm{up}$) for a given incoming infrared power $P_\mathrm{IR}$ (photon flux $\phi_\mathrm{IR}=P_\mathrm{IR}/h\nu_\mathrm{IR}$) and pump power $P_0$. The quantum efficiency follows from $\eta=\phi_\mathrm{up}/\phi_\mathrm{IR}$. 
One approach to this calculation was presented in Refs.~\cite{roelli2016,roelli2020}, treating the collection of $N$ molecules as a single collective molecular oscillator, with the optomechanical coupling rate scaling as $\sqrt{N}$. 
This approach can be used based on the results presented in Sec.~\ref{sec:EMsimulation} and \ref{sec:molecules} above, and yields conversion efficiencies of the same order of magnitude as measured according to Sec.~\ref{sec:calibration} (cf. Table~\ref{tab:statistics}), i.e. $10^{-12}$ to $10^{-11}$ from incoming IR photon to collected VIS photon. 
Below, we provide an alternative estimate following more closely the language and methods used in the Raman and IR spectroscopy communities, which further confirms the consistency of our model with our measurements.

We first perform the calculation for a single molecule, and then assume that $P_\mathrm{up}$ scales linearly with the number of molecules $N=A_\mathrm{cav}\rho$, where $A_\mathrm{cav}$ is the effective nanocavity area and $\rho$ the molecular surface density \cite{ahmed2021}.
Since the molecules occupy a region much smaller than the pump wavelength, the degree of spatial coherence is not expected to modify this scaling.
Moreover, we operate in a pump power regime where collective amplification is negligible \cite{zhang2020c}.
This approach idealises the overlap of IR and visible plasmonic modes; in reality, $N$ is an upper bound for the effective number of molecules participating in the upconversion. \\

The IR-driven phonon occupancy $n_\mathrm{coh}$ created in excess of the thermal occupancy $n_\mathrm{th}$ is the ratio of phonon creation rate to decay rate: $n_\mathrm{coh} = \frac{\gamma_+}{\gamma_-}$, where each rate is expressed as:
\[  \gamma_- = 1/\tau_\mathrm{m} ~~~\textrm{and} ~~~ \gamma_+ = \frac{1}{h\nu_\mathrm{IR}} \frac{P_\mathrm{IR}}{A_\mathrm{IR}}\sigma_\mathrm{IR} \]
Here, we have introduced the following quantities:
\begin{itemize}
    \item $\tau_\mathrm{m}$ is the vibrational $1/e$ energy decay time. This vibrational lifetime is half of the coherence time in the absence of pure dephasing \cite{velez2020}.
    \item $h\nu_\mathrm{IR}$ is the energy of incoming IR photons and corresponding vibrational quanta.
    \item $A_\mathrm{IR}$ is the effective IR spot area. We can link it to the numerical aperture for a diffraction limited spot by $A_\mathrm{IR}\simeq \pi D^2/4$ where $D$ is the spot FWHM, which at the wavelength $\lambda_\mathrm{IR}$ is approximately: $D\simeq \frac{\lambda_\mathrm{IR}}{2 NA}$.
    \item $\sigma_\mathrm{IR}$ is the IR absorption cross section (which has units of area).
\end{itemize}

Knowing $n_\mathrm{coh}$, we express the corresponding upconverted Raman power: 
$$ P_\mathrm{up} = \frac{P_0}{A_0} \sigma_\mathrm{R}  n_\mathrm{coh}$$
Here, we have introduced the following quantities:
\begin{itemize}
    \item $A_0$ is the effective pump laser spot area. As above, we can write $A_0\simeq \pi D^2/4$ with $D\simeq \frac{\lambda_0}{2 NA}$.
    \item $\sigma_\mathrm{R}$ is the total Raman cross section.
\end{itemize}

From this quantity, we can find the expression of the upconversion efficiency $\eta$ per unit of pump power (Unit: W$^{-1}$), considering $N$ molecules equally contributing to the upconversion, and integrating the upconverted power in the $4\pi$ solid angle:
$$ \tilde{\eta} = \frac{\phi_\mathrm{up}}{\phi_\mathrm{IR}}\frac{N}{P_0} = \frac{N \sigma_\mathrm{IR} \sigma_\mathrm{R} \, \tau_\mathrm{m}}{A_0 A_\mathrm{IR} \, h\nu_\mathrm{up}} $$
For an isotropic Raman scattered field, the conversion efficiency when taking into account the collection through the objective's solide angle will be reduced by a factor $(1-\sqrt{1-(NA)^2})/2$, where $NA$ is the numerical aperture (we consider the medium is air), which is 0.28 for NA=0.9 as in our experiment.


These expressions are valid for molecules in free space, and we take chemical enhancement into account in our molecular simulations (Sec.~\ref{sec:molecules}) by linking a gold atom to the sulfur atom of BPhT.
For a molecule coupled to a nanocavity, we expect the conversion efficiency to be enhanced by the factor $\tilde{F}=\beta_\mathrm{IR}^2 \beta_0^2 F_\mathrm{rad}$, where the $\beta$'s are field enhancement factors $|E_{loc}|/|E_{inc}|$ at the incoming IR and pump wavelengths, and $F_\mathrm{rad}$ is the radiative enhancement for a dipole placed inside the nanocavity, computed as explained in Section~\ref{sec:EMsimulation} and taking into account the solid angle collected by the objective.

\begin{figure}
    \centering
    \includegraphics[scale=1.2]{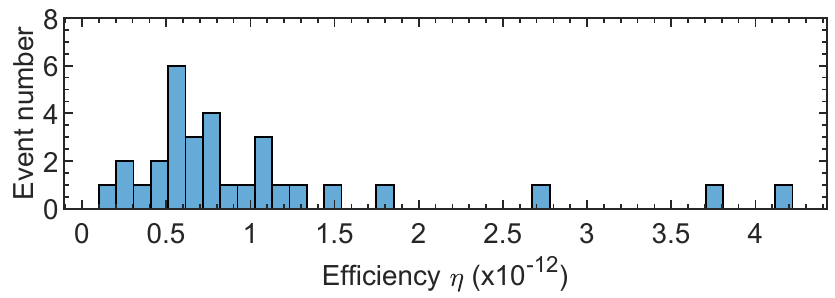}
    \caption{\textbf{Histogram of the conversion efficiency from Table~\ref{tab:parameters}.}
}
    \label{fig:eta_histogram}
\end{figure}

\begin{figure}
    \centering
    \includegraphics[scale=1]{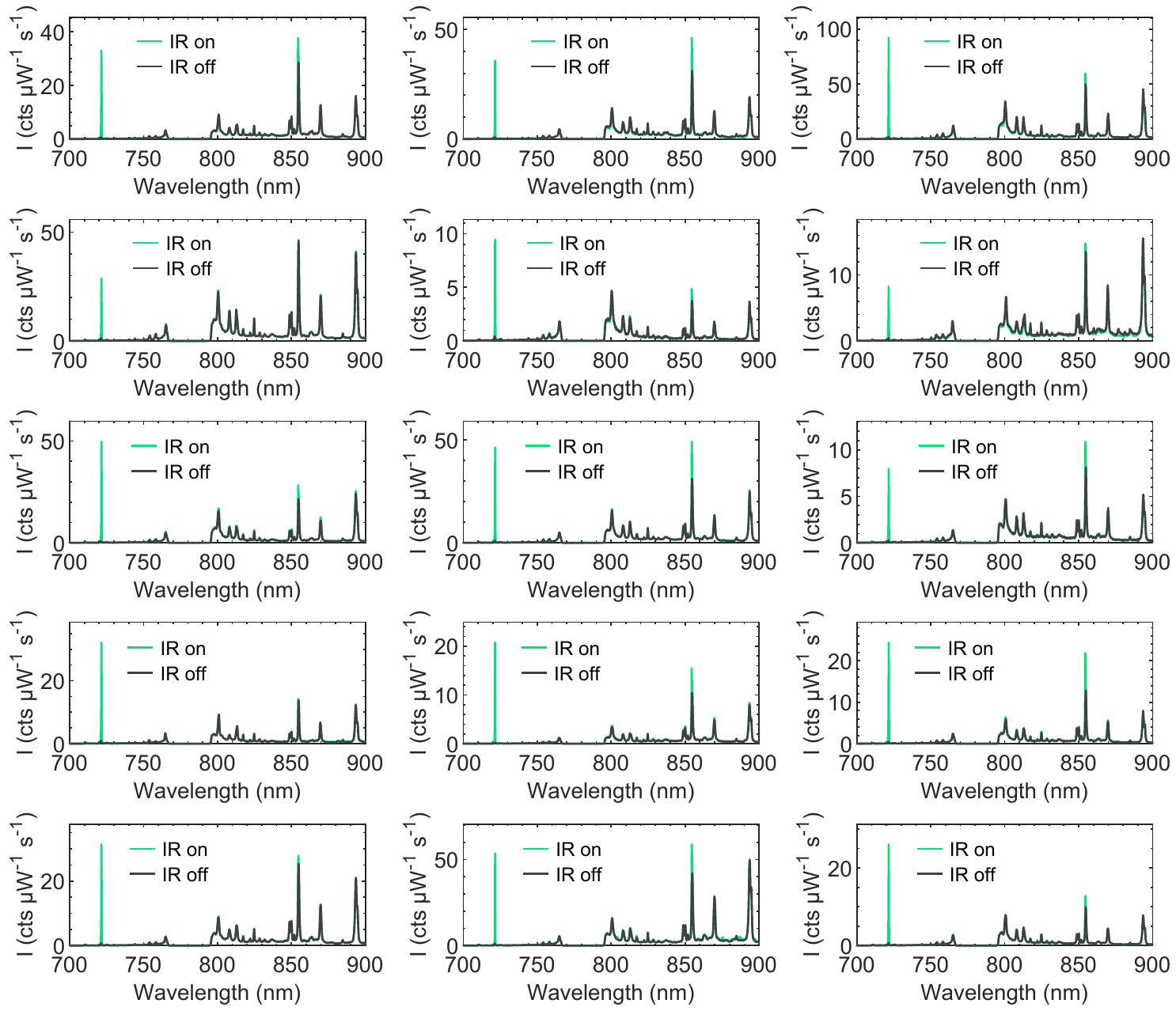}
    \caption{\textbf{More examples of upconversion spectra from different nanocavities.}
    The pump light is a TiSa laser at 785 nm with power of 10 $\mu$W. 
    IR power, 580 $\mu$W. Acquisition time, 10~s.}
    \label{fig:NPstatistics}
\end{figure}

\begin{figure}
    \centering
    \includegraphics[scale=1]{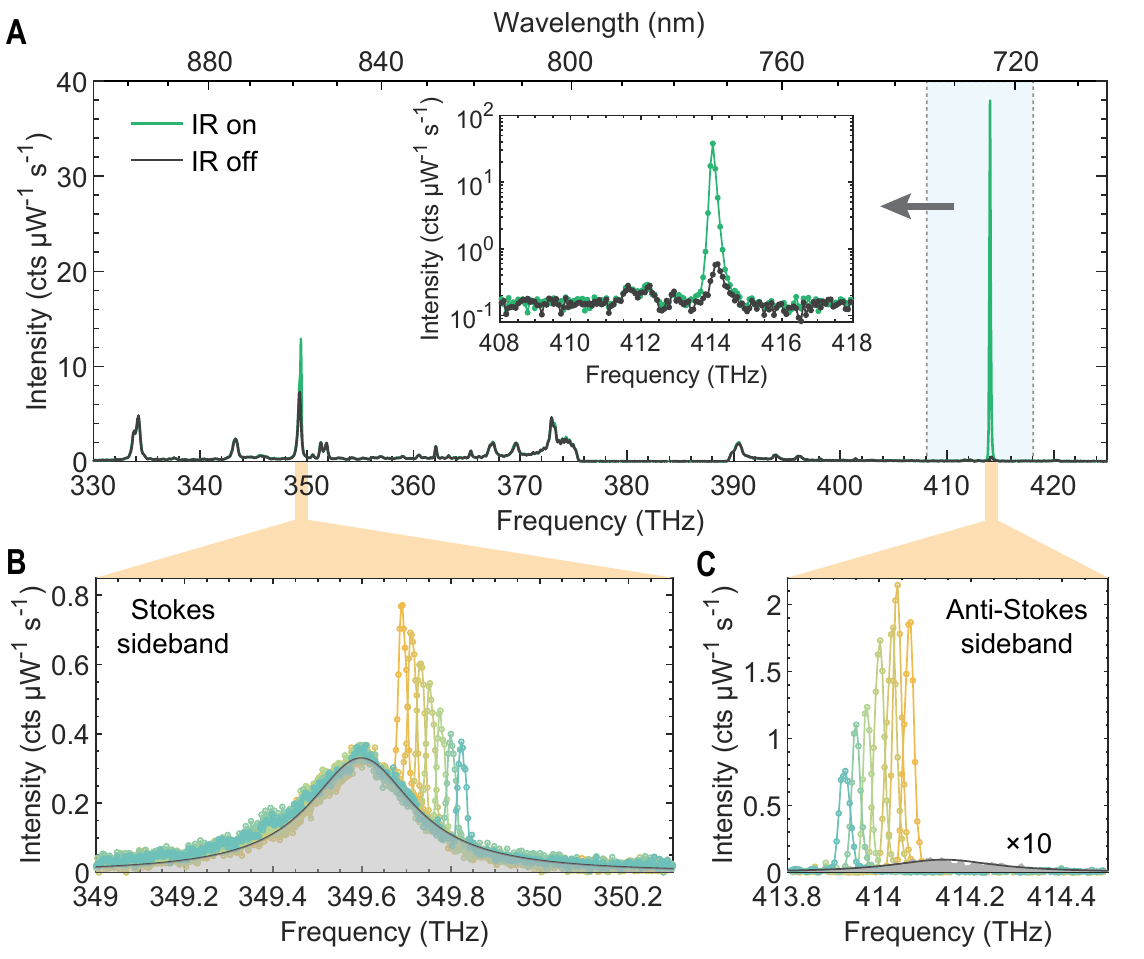}
    \caption{\textbf{IR frequency scanning upconversion measurement using a TiSa laser tuned at 785~nm as pump.} 
    (\textbf{A}) Low-resolution, full-range Raman spectra from a single nanocavity without (black line) and with (green line) incoming IR radiation (550 $\mu$W on the nanocavity), for 11~$\mu$W pump power at 785~nm on the sample. The inset shows a log scale view of the anti-Stokes sideband. Acquisition time, 10~s.
    (\textbf{B,C}) High-resolution Stokes (\textbf{B}) and anti-Stokes (\textbf{C}) spectra observed when tuning the signal beam with normalized IR radiation power of 135~$\mu$W. 785 nm pump power: 11~$\mu$W. The grey line in \textbf{C} and \textbf{D} are the spectra without IR radiation, where the grey areas are the corresponding Lorentzian fits. Acquisition time of the upconverted (spontaneous) Stokes and anti-Stokes spectra are 30~s (300~s) and 10~s (600~s), respectively.}
    \label{fig:TiSa785_QCLTune}
\end{figure}

\begin{figure}
    \centering
    \includegraphics[scale=1]{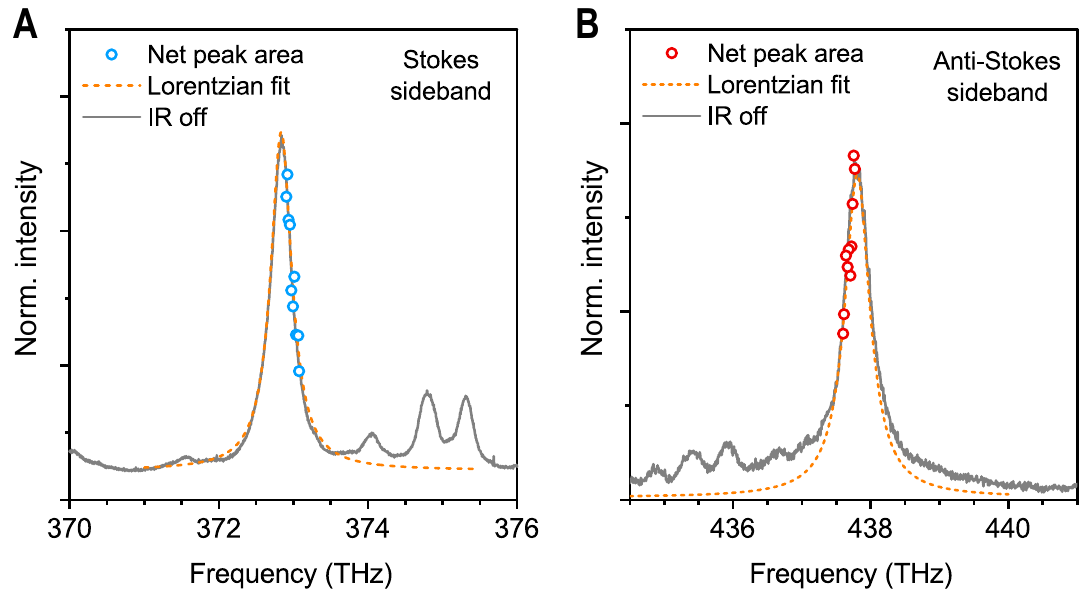}
    \caption{\textbf{IR frequency dependent upconverted signal intensity vs. spontaneous Raman spectrum.} 
    (\textbf{A,B}) Net peak area of the upconverted signal as the function of the peak frequency (obtained from Fig.~\ref{fig:spectrum}D,E by Lorentian fits) and its comparison with the spontaneous Raman spectra on the Stokes (\textbf{A}) and anti-Stokes (\textbf{B}) sidebands. The data are the same as in Fig.~\ref{fig:spectrum}D and E. The orange lines are the Lorentzian fits to the frequency dependent net peak area, where the peak position and linewidth for fitting are taken from the value of the corresponding spontaneous Raman peak. The intensities of the spontaneous Raman spectra were normalised to match the height of the orange curve for a direct comparison. The fitting profiles match well with the spontaneous Raman peak, suggesting that the upconversion intensity is determined by the detuning between IR signal and vibrational resonance. The effect of the plasmonic resonance is not observable due to its much broader width.
    These plots highlight an interesting feature: the spontenous Stokes and anti-Stokes Raman shift are not exactly the same. Such a small discrepancy would require very careful calibration of the spectrometer to be measured precisely; but the upconversion signal provides a perfect calibration tone. This observation suggests that our upconversion method could be used in ultra-high resolution dual-sideband Raman spectroscopy.}
    \label{fig:ProfileFit}
\end{figure}

\paragraph{Results of numerical estimates}
Table~\ref{tab:parameters} summarises the parameters extracted from molecular DFT and electromagnetic FEM simulations as explained in Sec.~\ref{sec:molecules} and \ref{sec:EMsimulation}, and corresponding to the experimental setup presented in the main text and in Sec.~\ref{sec:acquisition}. 
From these parameters, the expected nanocavity-enhanced upconversion efficiency is computed to be 
$$\eta = \tilde{\eta} \cdot \tilde{F} \simeq 6 \times 10^{-12}$$ in agreement with the experimental efficiency on the order of $10^{-12}$ measured for typical nanocavities (Table~\ref{tab:statistics} and Fig.~\ref{fig:eta_histogram}). 
In addition, the computed IR-driven phonon number for 600~$\mu$W IR power is $n_\mathrm{coh}\simeq 0.11$, also in agreement with typical values measured on the anti-Stokes sideband and reported in Table~\ref{tab:statistics}. 

\begin{table}[h!]
\centering
\begin{tabular}{||c|c|c|c|c|c|c|c||} 
 \hline
 Parameter & $\sigma_\mathrm{IR}$ & $\sigma_\mathrm{R}$ & $\rho$ & $A_\mathrm{cav}$ & $N$ & $\nu_\mathrm{IR}$ & $\tau_\mathrm{m}$ \\[1 ex] 
 Unit & cm$^2$ & cm$^2$ & cm$^{-2}$ & cm$^2$ & molecules & THz & ps \\ [1 ex]
 Value & $7\times 10^{-18}$ & $5\times 10^{-29}$ & $5\times 10^{14}$ & $1.6\times 10^{-12}$ & 785 & 32.4 & 0.53 \\ [1 ex]
 \hline
 \hline
 Parameter & $\beta_\mathrm{IR}$ & $\beta_0$ & $F_\mathrm{rad}$ & $\nu_\mathrm{up}$ &  &   &   \\[1 ex] 
 Unit & factor & factor & factor & THz &   &   &   \\ [1 ex]
 Value & 100 & 500 & $1.5\times 10^{4}$ & 437.5 &   &   &   \\
 \hline
\end{tabular}
\caption{Parameter values extracted from simulations and used in the theoretical efficiency estimate. The vibrational lifetime $\tau_\mathrm{m}$ is estimated from a spontaneous Raman linewidth of 300~GHz (see Fig.~\ref{fig:Fit_VS}).}
\label{tab:parameters}
\end{table}

\paragraph{Remark}
For a fair estimate of the enhancement factor, we should compare our upconversion efficiency to that of a hypothetical molecular monolayer on a flat gold substrate, for which all molecules overlapping with the pump beam would contribute to upconversion. 
In this case, $\tilde{F}$ defined above must be multiplied by $N/N_0 = A_n/A_L$, where $N_0$ is the number of molecules covering the pump laser spot area $A_L$, while $A_n$ is the effective area of the nanocavity containing the $N$ active molecules. 
With our experimental parameters, we find  $\tilde{F}\simeq 3\times 10^{13}$ and $A_n/A_L \simeq 10^{-3}$, so that our effective upconversion enhancement factor is in excess of $10^{10}$ using this estimate.



\section{Supplementary Text}
\subsection{Possible intepretations of the IR-driven vs. natural linewidths} \label{sec:linewidth}

In this section, we discuss the interpretation of the resolution-limited linewidth of the upconverted signal as seen in Fig.~\ref{fig:spectrum}D,E, and show that it most likely reflects a coherently displaced thermal state of the collective molecular oscillator.
Central to this discussion is the origin of the spontaneous Raman linewidth $\Delta\nu$ (on the order of 230 to 350~GHz as measured without IR drive), which is related to the vibrational decay time $\tau_\mathrm{m}$ by $\tau_\mathrm{m}=1/2\pi\Delta\nu$ and to the vibrational coherence time $\tau_\mathrm{m,coh}$ by $\tau_\mathrm{m,coh}=1/\pi\Delta\nu = 2\tau_\mathrm{m}$. 
Two extreme physical situations can \textit{a priori} be encountered (with any intermediate situation being possible):
\begin{enumerate}
    \item[(i)] The spectral linewidth $\Delta\nu$ of the spontaneous Raman scattered field is dominated by \textit{homogeneous broadening}, translating into $\tau_\mathrm{m,coh}\simeq 1$ to 1.4~ps for the coherence time of the collective vibrational mode. 
    \item[(ii)] The spontaneous linewidth is dominated by \textit{inhomogeneous broadening}, and each individual molecule has a much longer intrinsic coherence time -- possibly longer than 40~ps, which would correspond to $\Delta\nu< 7.5$~GHz as observed for the upconversion peak.
\end{enumerate}
In case (i), the resolution-limited, sub-10~GHz linewidth of the upconverted (i.e. IR driven) Raman peak would find its explanation in the coherent nature of optomechanical transduction, which results in a coherently displaced thermal state of the oscillator.
In case (ii), even if upconversion were fully incoherent, the resonant driving of a sub-ensemble of molecules could yield data similar to Fig.~\ref{fig:spectrum}. 

\paragraph{Discussion on coherence lifetime} 
As mentioned above, case (ii) would entail single molecule vibrational coherence times in excess of 40~ps. 
For molecules bound to a metal and at ambient temperature, this coherence time appears improbably long. 
For example, the linewidth of a specific Raman peak of individual Nile Blue molecules adsorbed on Ag surfaces was measured to be 3 to 3.5~cm$^{-1}$ at room-temperature \cite{artur2011}, corresponding to a coherence time on the order of 3~ps. 

The only report we know of in the literature that could support much longer coherence times is Ref.~\cite{yampolsky2014}, where the authors studied \textit{trans}-1,2-bis-(4-pyridyl) ethylene (BPE) molecules embedded inside gold nanoparticle dimers encapsulated in porous silica shells.
In  Fig.~1d of Ref.~\cite{yampolsky2014} the continuous-wave Raman spectrum appears Gaussian, in agreement with its being dominated by inhomogeneous broadening. 
When performing time-resolved, surface-enhanced coherent anti-Stokes Raman spectroscopy (tr-SECARS), the authors occasionally found that the coherent beat signal between two nearby vibrational modes persisted for at least 10~ps, which they interpreted as instances of single-molecule signal free of inhomogenous broadening.
We note, however, that the $>10$~ps persistence of the beat note in tr-SECARS does not imply that the continuous-wave Raman linewidth of a single molecule would be observed as smaller than 30~GHz. 
Indeed, the beat note in tr-SECARS is only sensitive to the relative phase between the two nearby vibrational mode, so that all common-noise is rejected. 
In particular, this measurement is not sensitive to common jitter of both Raman peaks. 
Finally, the molecules used in Ref.~\cite{yampolsky2014} are not bound to gold by thiol groups, as in our experiment. 
Thiol bounds are strong and modify the Raman and IR polarisability of the molecules, thereby most likely affecting also there vibrational lifetimes.

\begin{figure}
    \centering
    \includegraphics[scale=1]{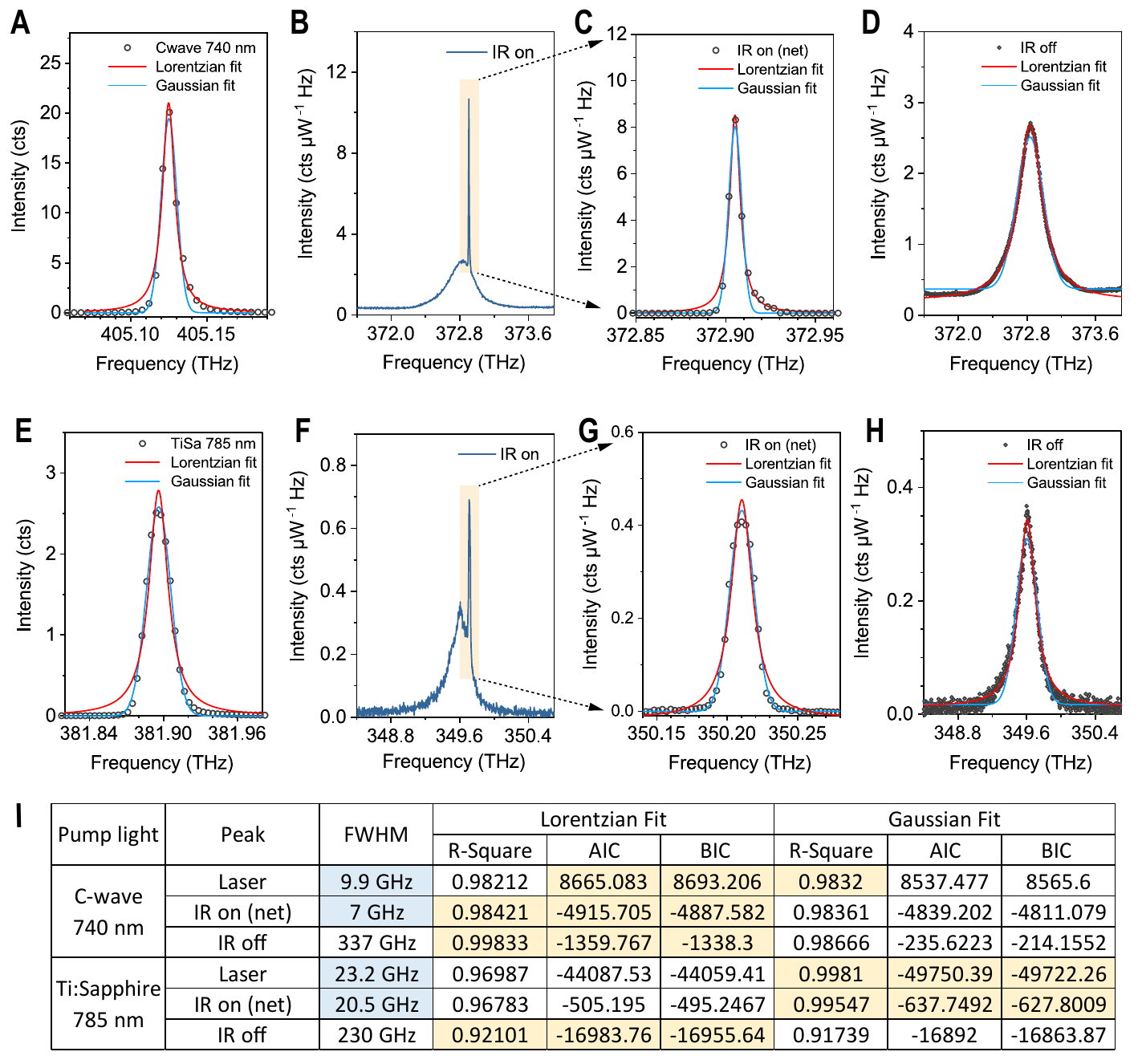}
    \caption{\textbf{Peak shape analysis.} 
    (\textbf{A-H}) Typical spectra of the laser (\textbf{A, E}), Stokes peak under IR radiation (\textbf{B, F}), net upconverted signal (spontaneous peak substracted, \textbf{C, G}) and the Stokes peak without IR radiation (\textbf{D, H}), along with their Lorentzian and Gaussian fits. Spectra of \textbf{B, D} and \textbf{F, H} are taken from Fig.~\ref{fig:spectrum}D and E and Fig.~\ref{fig:TiSa785_QCLTune}B and C, respectively. (\textbf{I}) Summary of the fit results from \textbf{A-H}, where R-square, Akaike's information criterion (AIC) and Bayesian information criterion (BIC) are used to identify which of Lorenzian or Gaussian function fits best with the peak. For R-square the higher value the better, and for AIC and BIC the lower value the better (BIC: difference greater than 10 gives decisive conclusion), which are labelled with yellow shadow in the table.
    Importantly, the spontaneous (IR off) lineshapes are better reproduced by Lorentzian functions, suggesting that homogeneous broadening significantly contributes to the linewidth.}
    \label{fig:Fit_VS}
\end{figure}

\begin{figure}
    \centering
    \includegraphics[scale=1]{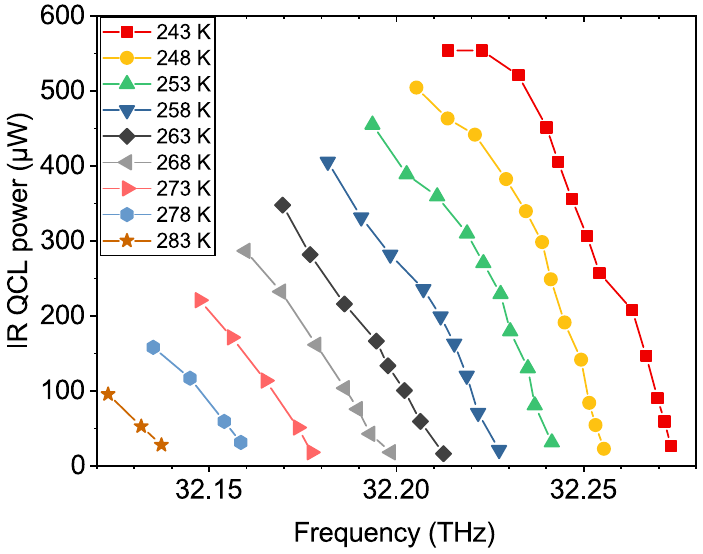}
    \caption{\textbf{QCL performance as probed by the upconversion spectra.}
    Power and frequency of the IR radiation from QCL are dependent on its working temperature and current, which can be probed in the optical domain using the high resolution upconversion spectrum (Fig.~\ref{fig:spectrum}D). The figure shows frequency versus power (on the nanocavity) of the QCL under different temperatures while scanning the current, which agrees well with the FTIR results provided from the QCL supplier. This result was used as a reference for the IR frequency tuning (Fig.~\ref{fig:spectrum}D and E) and IR power dependence (Fig.~\ref{fig:power}B).}
    \label{fig:QCLpeformance}
\end{figure}

\begin{figure}
    \centering
    \includegraphics[scale=0.9]{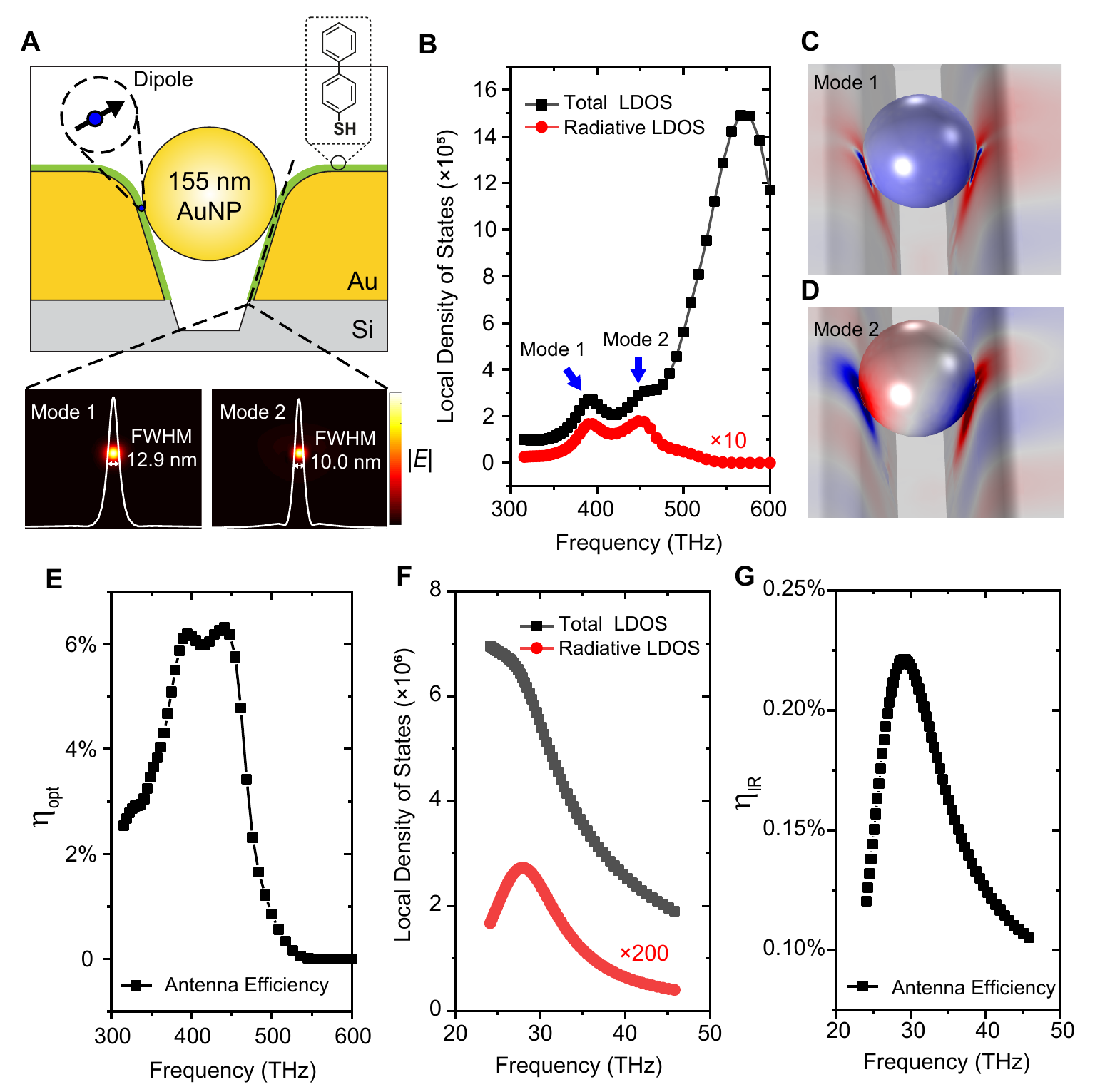}
    \caption{\textbf{Simulated electromagnetic response of the nanocavity in the visible and infrared ranges}.
    The local density of states (LDOS) enhancement (\textbf{B, F}) and antenna efficiency (\textbf{E, G}) of the modes in visible (\textbf{B, E}) and infrared range (\textbf{F, G}), respectively, according to the simulation schematics (\textbf{A}) where an electric dipole was placed in the nanogap. The electric field distribution of two modes on a slice parallel to the nanogroove's sidewall were plotted (lower panel of \textbf{A}), whose mode confinement were demonstrated by their full width at half maximum (FWHM).  \textbf{C, D} are the charge distribution of two modes near 392.3 THz and 449.5 THz shown in \textbf{B}.}
    \label{fig:simulation}
\end{figure}

\begin{figure}
    \centering
    \includegraphics[scale=0.75]{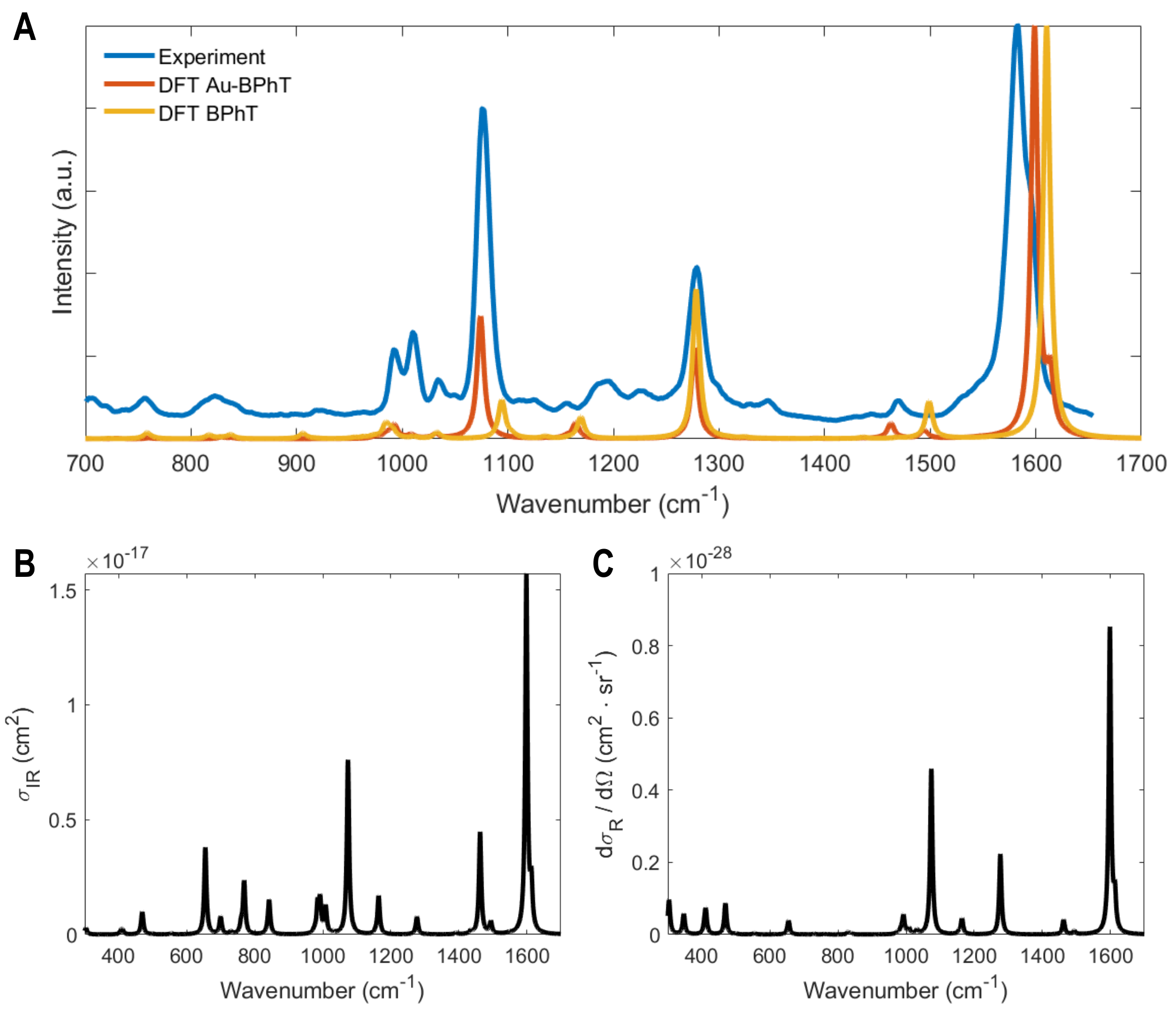}
    \caption{\textbf{Simulated response of the molecule in the visible and IR domains}.
    (\textbf{A}) Comparison of experimental Raman spectra with DFT simulations of Au-BPhT and BPhT molecules. 
    (\textbf{B}) Absorption cross-section and (\textbf{C}) differential Raman cross-section spectra of a single Au-BPhT molecule, with the same respective orientation to the external field as in panel \textbf{A}.}
    \label{fig:dft}
\end{figure}

\paragraph{Discussion on spontaneous Raman lineshape}
The most convincing and direct evidence we have in support of Case (i) (linewidth dominated by homogeneous broadening) is a careful analysis of the peak lineshape, as summarised in Fig.~\ref{fig:Fit_VS}.
More specifically, Figs.~\ref{fig:Fit_VS}D and H show two typical spontaneous Stokes peaks and their best fits (in the sense of least squared error) with Lorentzian (red lines) and Gaussian (blue lines) functions. 
Visual inspection close to the center and the edges of the peak favor the Lorentzian fit as more faithful. 
Note that the baseline is skewed by the presence of nearby Raman peaks, in particular at higher frequencies (i.e. smaller Raman shifts for the Stokes sideband). 
To be more quantitative, we conducted two statistical tests: Akaike's information criterion (AIC) and Bayesian information criterion (BIC) to elect the best of the two fits. 
Both tests favor the Lorentzian fit, Fig.~\ref{fig:Fit_VS}. 

We conclude that the observed spontaneous linewidth is more likely dominated by homogeneous broadening.
Consequently, the much narrower linewidth of the upconversion peak supports the coherent nature of the upconversion process. 
Nevertheless, we argue that it would be interesting to conduct future experiments to better estimated the extent of inhomogeneous broadening and to quantify the degree of coherence of the upconverted signal with respect to the incoming IR signal.


\section{Supplementary Figures and Tables}

\newpage
\begin{table}
    \centering
    \includegraphics[scale=1.3]{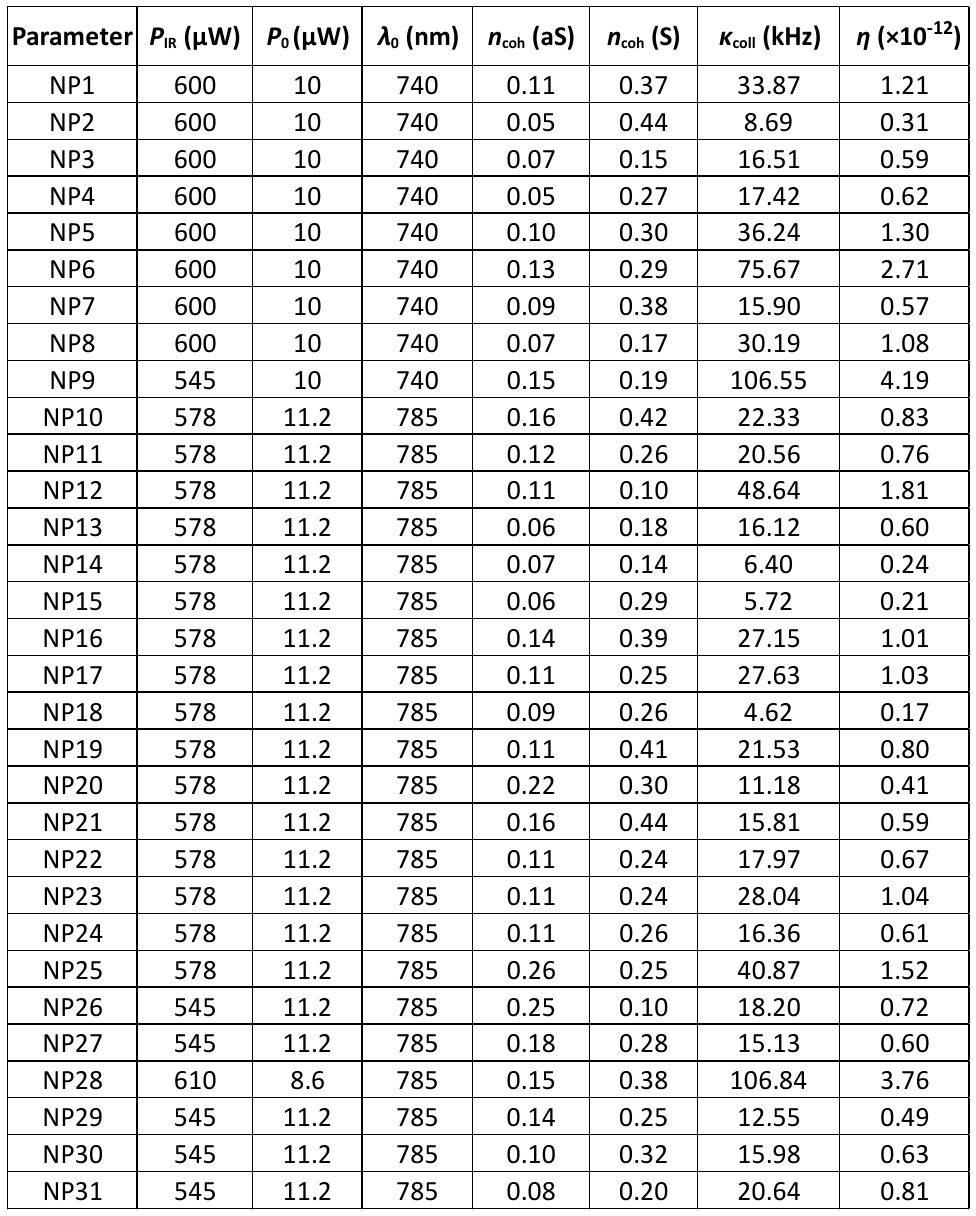}
    \caption{\textbf{Summary of experimental results on several nanocavities.}
$\kappa_\mathrm{coll}$ is the net upconverted signal collected by the objective on the anti-Stokes sideband, which is estimated as described in Section \ref{sec:calibration}. The experimental end-to-end upconversion efficiency is then computed as $\eta = \kappa_\mathrm{coll}h\nu_{\mathrm{IR}}/P_{\mathrm{IR}}$, where $h$ the Planck's constant, $\nu_{\mathrm{IR}}$ the IR radiation frequency.
}
    \label{tab:statistics}
\end{table}

\end{document}